# Statistical switching kinetics in ferroelectrics

X.J. Lou[*]

Department of Materials Science and Engineering, National University of Singapore, 117574, Singapore

**Abstract**:

By assuming a more realistic nucleation and polarization reversal scenario we build a new statistical switching model for ferroelectrics, which is different from either the Kolmogorov-Avrami-Ishibashi (KAI) model or the Nucleation-Limited-Switching (NLS) model. After incorporating a time-dependent depolarization field this model gives a good description about the retardation behavior in polycrystalline thin films at medium or low fields, which can not be described by the traditional KAI model. This model predicts correctly $n=1$ for polycrystalline thin films at high $E_{appl}$ or ceramic bulks in the *ideal* case.

The traditional method to describe the switching kinetics in ferroelectrics is the KAI model [1], based on the classical theory by Kolmogorov [2] and Avrami [3]. For a fully poled ferroelectric capacitor driven by an applied field $E_{appl}$, the KAI theory gives the polarization change $\Delta P(t)$ to be:

$$\frac{\Delta P(t)}{2P_s} = 1 - \exp[-(t/\tau)^n] \qquad (1)$$

---

[*] Correspondence email: mselx@nus.edu.sg

where $n$ and $\tau$ are the effective dimensionality and characteristic time, respectively. Although the KAI model, which assumes an infinite crystal as well as unrestricted domain growth, has been successfully used to describe the switching kinetics in single-crystalline [4] as well as epitaxial thin-film ferroelectrics [5], it confronts problems in correctly describing the domain reversal behavior in polycrystalline ferroelectric thin films, particularly at low applied fields [6-8]. These behaviors have been explained by polarization processes with a broad distribution of relaxation times [6], a NLS model [8] or the Lorentzian distribution of logarithmic domain-switching times very recently [7]. Furthermore, the effect of the depolarization field $E_{dep}$ during polarization switching has been totally ignored in these models and scenarios. Although it is a good approximation for bulk materials, it fails for thin-film samples, in which it is apparent that the depolarization effect plays an important role in determining their switching behaviors, especially at medium and low fields.

First of all, let us divide the total area of a ferroelectric capacitor *uniformly* into $M_0$ parts. Then we assume that under the total field (i.e., the sum of $E_{appl}$ and $E_{dep}$) polarization switching in this capacitor takes place in a part-by-part or region-by-region manner due to the *blocking* effect of grain boundaries [9] and/or 90 ° domain walls [10], consistent with the microscopic observations in ferroelectric polycrystalline thin films [9-11]. In other words, polarization reversal in each part of the film occurs *independently* by formation of an opposite nucleus, followed by forward and sideways growth of opposite domain. The nucleation effect at the edge of reversed parts on its neighboring parts (i.e. the effect of domain wall motion crossing the boundaries between one part and another) is neglected here and will be considered elsewhere [12].

The depolarization field in a poled Pt/PZT/Pt capacitor with interface layers has been written as [13]:

$$E_{dep}(t) = \frac{d_i P(t)}{d \varepsilon_i \varepsilon_0} \qquad (2)$$

where $d$ and $d_i$ are the thickness of the film and the interface layer, respectively. $\varepsilon_i$ is the interface dielectric constant. $P(t)$ is the time-dependent polarization. Assuming $d$=200 nm, $d_i$=2 nm, $\varepsilon_i$=40 [14] and $P(t)$=30 $\mu$C/cm$^2$, we have $E_{dep}(t)$~85 kV/cm, which is indeed large enough to play some role during switching process in ferroelectric thin films, especially at low fields. Note that the depolarization field *unavoidably* appears to some extent in insulating thin-film ferroelectrics due to the poor screening of the bound charge at the interface induced by an interface passive layer and/or polarization gradient near the electrode and/or a finite electrode screening length [15]. In general, it can be written as:

$$E_{dep}(t) = \beta \frac{P(t)}{\varepsilon_f \varepsilon_0} \qquad (3)$$

where $\beta$ is the depolarization factor, $\varepsilon_f$ is the ferroelectric dielectric constant. (Note that for the convenience of mathematical estimation we will use Eq (2) rather than Eq (3) in the following derivation. However, replacing Eq (2) by the general form Eq (3) or any other specific form for $E_{dep}$ is straightforward.) So, the total field experienced by the film is:

$$E_{tot}(t) = E_{appl} + E_{dep}(t) = E_{appl} + \frac{d_i P(t)}{d \varepsilon_i \varepsilon_0} \qquad (4)$$

It can be seen that the total field is not a constant as assumed in other models, but a time-dependent quantity [16]. Let's define the direction of $E_{appl}$ to be positive. Since $E_{dep}(t)$ is always antiparallel to $P(t)$, $E_{dep}(t) = \frac{d_i P(t)}{d \varepsilon_i \varepsilon_0} > 0$ at earlier switching stage, but $E_{dep}(t) = \frac{d_i P(t)}{d \varepsilon_i \varepsilon_0} < 0$ at later switching stage, depending on whether the total $P(t)$ changes its sign.

Let us begin with a fully poled ferroelectric thin-film capacitor with polarization $P_{M_0}$ at $t$=0~10$^{-13}$ s (the period of an optical phonon) when the poling field is just removed. For simplicity we imagine that the external field $E_{appl}$ is *immediately* applied without the relaxation (or retention loss) of $P_{M_0}$ induced by the depolarization field (i.e. the waiting time $t_w$ is slightly larger than 10$^{-13}$ s, see inset in Fig 1). We will return to the situation in which there are noticeable time intervals (therefore polarization relaxation) between the pulses (i.e., $t_w$>>10$^{-13}$ s). Then we assume $1/\xi(t_N)$ ($\xi$>>1) to be the probability that one more part of the total area switches after $t_c$ from the time point where $N$ parts have *just* switched. $t_N$ is defined as the time interval that the $N$th part takes to switch. $t_c$ is a characteristic time and can be chosen arbitrarily as long as it ensures that $1/\xi(t_N)$<<1 for any $t_N$, and it disappears later on as we will see. So the probability that one part will not switch after $t_c$ from time $t_0$ is [1-1/$\xi(t_0)$]. Then the probability that one part will survive from switching after $t_1$ ($t_1$>>$t_c$) from time $t_0$ is $\left(1 - \frac{1}{\xi(t_0)}\right)^{t_1/t_c}$. According to the definition of $t_N$, the total number of the parts that retain their polarization after $t_1$ is:

$$M_0 - 1 = M_0 \left(1 - \frac{1}{\xi(t_0)}\right)^{t_1/t_c} \qquad (5)$$

which can be rearranged into:

$$\frac{M_0 - 1}{M_0} = \left(1 - \frac{1}{\xi(t_0)}\right)^{t_1/t_c} \qquad (6)$$

Taking natural logarithm on both sides of Eq (6), we get:

$$\ln \frac{M_0 - 1}{M_0} = \frac{t_1}{t_c} \ln \left( 1 - \frac{1}{\xi(t_0)} \right) = \frac{t_1}{t_c} \left[ -\frac{1}{\xi(t_0)} - \frac{1}{2} \left( \frac{1}{\xi^2(t_0)} \right) + \cdots \right] \qquad (7)$$

Considering that $\xi(t_0) \gg 1$, we can neglected all the higher-order terms and have:

$$\frac{M_0 - 1}{M_0} = exp \left( -\frac{t_1}{t_c \xi(t_0)} \right) \qquad (8)$$

Recall Merz's law [17]:

$$\frac{1}{t_{sw}} = \frac{1}{t_\infty} \cdot exp \left( -\frac{\alpha}{E_{tot}} \right) \qquad (9)$$

where $t_{sw}$ and $t_\infty$ is the switching time for $E_{tot}$ and an infinite field, respectively. $\alpha$ is the activation field for switching. According to the meaning of $1/\xi(t_0)$ defined above, we have:

$$\frac{1}{\xi(t_0)} = \frac{t_c}{t_{sw}} = \frac{t_c}{t_\infty} \cdot exp \left( -\frac{\alpha}{E_{tot}(t_0)} \right) \qquad (10)$$

Inserting Eq (4) and Eq (10) into Eq (8), we have:

$$\frac{M_0 - 1}{M_0} = exp \left( -\frac{t_1}{t_\infty} \cdot exp \left( -\frac{\alpha}{\left( E_{appl} + \frac{d_i P_{M_0}}{d \varepsilon_i \varepsilon_0} \right)} \right) \right) \qquad (11)$$

Similarly, we get a series of *feedback* equations:

$$\frac{M_0 - 2}{M_0 - 1} = exp \left( -\frac{t_2}{t_\infty} \cdot exp \left( -\frac{\alpha}{\left( E_{appl} + \frac{d_i P_{M_0 - 1}}{d \varepsilon_i \varepsilon_0} \right)} \right) \right)$$

$$\vdots$$

$$\frac{M_0 - N - 1}{M_0 - N} = exp\left(-\frac{t_{N+1}}{t_\infty} \cdot exp\left(-\frac{\alpha}{(E_{appl} + \frac{d_i P_{M_0-N}}{d\varepsilon_i \varepsilon_0})}\right)\right) \qquad (12)$$

$$\vdots$$

$$\frac{M_0 - M_0}{M_0 - M_0 + 1} = \frac{0}{1} = exp\left(-\frac{t_{M_0}}{t_\infty} \cdot exp\left(-\frac{\alpha}{(E_{appl} + \frac{d_i P_1}{d\varepsilon_i \varepsilon_0})}\right)\right)$$

where $N = [0, 1, \cdots, (M_0 - 1)]$ and $P_{M_0-N} = \frac{M_0 - 2N}{M_0} P_{M_0}$. $P_{M_0-N} > 0$ when $N < M_0/2$; $P_{M_0-N} \leq 0$ when $N \geq M_0/2$. Note that Eq (12) contains *no* adjustable parameter. $t_\infty$ can be worked out both experimentally and theoretically for a specific sample. Merz obtained $t_\infty$=10 ns for a BTO single-crystal of 20 μm ($t_\infty$ decreases as $d$ decreases) [17], in agreement with $t_\infty$=13 ns achieved by Scott *et al*. in submicron PZT films [18]. Notice that the fastest switching time measured so far is 220 ps in a circuit with small *RC* constant of ~ 45 ps [19], where $t_\infty$ should be around ~100 ps or less. From a theoretical point of view, we could estimate $t_\infty \sim t_{pg}$=$d/v$, where $t_{pg}$ is the propagation time of needlelike domains, $v$ is the sound velocity (~2000 m/s). For a film with $d$=200 nm, we get $t_\infty \sim t_{pg}$=100 ps. One can see that $t_\infty$ depends very much on the parameters of a specific sample. $t_\infty$ varies from 100 ps to 10 ns for normal thin films. For simplicity, we use $t_\infty$=1 ns in this work.

Let's make some remarks about this model and give some predictions using it.

(1) *$M_0$* **effect**.

The $(Y, X)$ plots of $(\frac{\Delta P(t)}{2P_{M_0}} = \frac{1 - \frac{P_{M_0 - N}}{P_{M_0}}}{2} = \frac{N}{M_0}, t = \sum_{i=1}^{N} t_i)$ have been plotted in Fig 1 according to Eq (12) for $M_0$=100, 500 and 1000, where we took $E_{appl}$=500 kV/cm, $P_{M_0}$=30 μC/cm², $d$=200 nm, $\varepsilon_i/d_i$=20 nm⁻¹, $\alpha$=500 kV/cm [5, 19-21], and $t_\infty$=1 ns as justified above. One can see that the change of $M_0$ has little effect on the curves in Fig 1. The profiles are stabilized when $M_0 \rightarrow \infty$ (e.g. the curve for $M_0$=500 essentially overlaps with that for $M_0$=1000). In view of the derivation of Eq (12) it is obvious that *there is no analytical equation that can describe the whole curve*; each data point has to be calculated using Eq (12) and $t = \sum_{i=1}^{N} t_i$ for given $\frac{\Delta P(t)}{2P_{M_0}} = \frac{N}{M_0}$. From Eq (12) one can see that the effect of the change of $t_\infty$ is to just slightly shift the whole curve along the $X$ axis without changing its profile.

## (2) Polycrystalline thin films at low or medium $E_{appl}$

The $(Y, X)$ plots of $(\frac{\Delta P(t)}{2P_{M_0}} = \frac{N}{M_0}, t = \sum_{i=1}^{N} t_i)$ have been plotted in Fig 2 according to Eq (12) for $E_{appl}$=(500, 400, 300, 200, 150, 100, 90, 70, 50, 30, 10, and 0) kV/cm, respectively. From Fig 2 we see that the curves shift to the right along the time axis for lower fields. We also see that at low $E_{appl}$ the switching process is highly retarded and expands over many time decades: it covers ~3 decades for $E_{appl}$=500 and 400 kV/cm, ~4 decades for $E_{appl}$=300 and 200 kV/cm, and ~6 decades for $E_{appl}$=150 kV/cm; it dramatically increases to ~15 decades for $E_{appl}$=100 kV/cm when $E_{appl}$ is comparable with the maximum $E_{dep}$~85 kV/cm; and the switching time $t_{sw}$ essentially goes to infinity for $E_{appl} < E_{dep}^{max}$ ~85 kV/cm (i.e, it can never be fully switched as expected) in our imaginary system. Note that the "fan structure" of the profiles and the retardation at low $E_{appl}$ are in good agreement with the observations in the literature on other polycrystalline thin-film samples: see Fig 4 in the work by Lohse *et al.* [6], Fig 2 in the work by Tagantsev et al. [8], Fig 1 in the work by Jo et al. [7] and Fig 2 in another work by Jo et al. [22]. Also note that the $E_{appl}$~0 kV/cm case essentially corresponds to the switching kinetics driven purely by the depolarization field (i.e. the case of retention loss) and will be discussed elsewhere [12]. Finally, it should be noted that the set-ups for switching measurements used by other people are different from the ideal one we assumed in this work (see inset in Fig 1): we used $t_w$~$10^{-12}$ s as aforementioned to avoid the polarization relaxation loss; Jo et al. used $t_w$~500 ns=5x$10^{-7}$ s [7, 22]; Tagantsev et al. used $t_w$~1 s [8]; and So et al. used 10 s [5]. It is well known that significant retention loss occurs within $10^{-3}$ s after the poling pulse is removed [15] (readers can also get some ideas about the magnitude of the loss from the curve of $E_{appl}$~0 kV/cm in Fig 2). Therefore, the real situation could be very complex: retention loss (or backswitching) driven by pure $E_{dep}$ takes place during time interval $t_w$ between pulse 2 and pulse 3 (inset in Fig 1), switching occurs driven by residual $E_{dep}+E_{appl}$ during time interval $t$ of switching pulse 3, and another retention loss in opposite direction takes place driven by opposite $E_{dep}$ during the time interval $t_w$ between switching pulse 3 and read pulse 4. So, instead of $2P_s = 2P_{M_0}$ used by this work, the researchers were essentially working on $2P_s = 2P_{M_0-N_R}$ , where $N_R$ is the number of relaxed or backswitched parts, and depends on the value of $t_w$. Therefore, using an as-short-as-possible $t_w$ is strongly recommended in order to simplify switching studies.

Note that despite this complication, the real situation could still be studied using the present theory by considering the loss effect during $t_w$.

(3) **Polycrystalline thin films at high $E_{appl}$ or polycrystalline/ceramic bulks**

If $E_{appl} \gg E_{dep}$ (i.e. $E_{dep}$ can be neglected), according to Eq (12) we have:

$$t_{N+1} = \frac{-t_\infty \, ln\left(\frac{M_0 - N - 1}{M_0 - N}\right)}{exp\left(-\frac{\alpha}{E_{appl}}\right)} \qquad (13)$$

So:

$$t = \sum_{i=1}^{N} t_i = \frac{-t_\infty}{exp\left(-\frac{\alpha}{E_{appl}}\right)} \left[ ln\left(\frac{M_0 - 1}{M_0}\right) + \cdots + ln\left(\frac{M_0 - N}{M_0 - N - 1}\right) \right]$$

$$= \frac{-t_\infty}{exp\left(-\frac{\alpha}{E_{appl}}\right)} ln\left(\frac{M_0 - N}{M_0}\right) \qquad (14)$$

So we have:

$$\frac{\Delta P(t)}{2P_{M_0}} = \frac{N}{M_0} = 1 - exp\left(-\frac{t}{t_\infty} \cdot exp\left(-\frac{\alpha}{E_{appl}}\right)\right) \qquad (15)$$

One can see that Eq (15), a simplified version of Eq (12) when $E_{appl} \gg E_{dep}$, is actually equivalent to Eq (1) when $n$=1 and assuming:

$$\frac{1}{\tau} = \frac{1}{t_\infty} exp\left(-\frac{\alpha}{E_{appl}}\right) \qquad (16)$$

i.e. Merz's law Eq (9). Fig 3 shows the profiles of $\Delta P(t)/2P_{M_0}$ as a function of $t$ according to Eq (15), where $E_{appl}$=(500, 400, 300, 200, 150, and 100) kV/cm. For comparison, the curves according to Eq (12) for these $E_{appl}$ have also been plotted. One can see that the deviation between the curve according to Eq (15) and the one according to Eq (12) decreases as $E_{appl}$ increases. At $E_{appl}$=500 kV/cm (where $E_{appl} \gg E_{dep}$), they almost overlap with each other, which justifies that $E_{dep}$ indeed can be neglected and Eq (15) holds for polycrystalline thin films at high fields. At medium or low $E_{appl}$, however, the effect of $E_{dep}$ is to promote switching at earlier switching stage and retard it at later switching stage (Fig 3).

Therefore, our model predicts that $n\sim1$ *ideally* for the switching curve in polycrystalline thin films at high $E_{appl}$, in good agreement with the data by Jo *et al*., who obtained $n_{app}=1\pm0.4$ at room temperature ($n_{app}$ denotes "apparent" $n$), $n_{app}=1\pm0.4$ at 150 K, $n_{app}=1\pm0.6$ at 80 K, and $n_{app}=0.9\pm0.6$ at 25 and 15 K for various fields (see Fig 1 and 2 in Ref [7].). So $n$ indeed centers at 1.The reason why some "apparent" $n$ values at higher $E$ deviate from the ideal value and slightly larger than 1 is because of the effect of domain-wall motion crossing boundaries of parts [12]. The reason why $n_{app}<1$ at low fields [7] is caused by the promotion (or retardation) effect of $E_{dep}$ at the earlier (or later) switching stage discussed above. Actually, an easy way to estimate the $n$ value in the KAI model (Eq (1)) is to see how many decades the curves in Fig 2 expand, e.g. one-decade expansion leads to $n_{app}\sim3$, 1.5 decades to $n_{app}\sim2$, $\sim3$ decades to $n_{app}\sim1$. Expansion over more than 4 decades always gives rise to $n_{app}<1$.

Our model (Fig 2 and Fig 3) also implies that the KAI model (which completely ignores $E_{dep}$) can not give good description for the data at low $E_{appl}$. Tentative fitting always gives very poor fitting quality together with $n_{app}<1$ (see Fig 1 and 2 in Jo's work [7].), and that is the reason why we call it "apparent" $n$, or $n_{app}$, here. Note that a $n$ value less than 1 is not physically reasonable according to the KAI model, since the growth dimensionality could never be less than 1 no matter how we mix the α model with the β model.

Finally let's look at the case of ceramic bulks where $E_{dep}$ could be neglected according to Eq (2). Polarization switching data by Verdier *et al*. (Fig 1 in Ref [23]) on commercialized virgin PZT bulk samples have been fitted using Eq (15) in Fig 4. One can see that Eq (15) having $n\sim1$ in the ideal case, derived by neglecting $E_{dep}$, can indeed give good description for the switching kinetics in polycrystalline/ceramic bulk materials. Best fitting (not shown) gives $n_{app}$=0.90, close to 1. Again, a weak retardation at later stage of the switching curve could also be seen due to the depolarization effect inside the material and/or near the electrode. The fitting curve in Fig 4 gives $\frac{1}{\tau} = \frac{1}{t_\infty} exp\left(-\frac{\alpha}{E_{appl}}\right) = 59250.17$. Since $E_{appl}$=20 kV/cm [23], we have $t_\infty = 6.21 \times 10^{-6}\ s$ for $\alpha$=20 kV/cm, $t_\infty = 1.39 \times 10^{-6}\ s$ for $\alpha$=50 kV/cm, and $t_\infty = 1.14 \times 10^{-7}\ s = 114\ ns$ for $\alpha$=100 kV/cm. For reasonable $\alpha$ values, $t_\infty$ in ceramic bulks is about 2 to 4 orders of magnitude higher than those in thin films, which is expected.

Figure Captions:

Fig 1 (color) the (Y, X) plots of $(\frac{\Delta P(t)}{2P_{M_0}} = \frac{1 - \frac{P_{M_0} - N}{P_{M_0}}}{2} = \frac{N}{M_0}, \sum_{i=1}^{N} t_i)$ according to Eq (12) for

$M_0$=100, 500 and 1000.

Fig 2 (color) the (Y, X) plots of $(\frac{\Delta P(t)}{2P_{M_0}} = \frac{N}{M_0}, t = \sum_{i=1}^{N} t_i)$ according to Eq (12) ($M_0$=1000) for

$E_{appl}$=(500, 400, 300, 200, 150, 100, 90, 70, 50, 30, 10, and 0) kV/cm.

Fig 3 (color) the profiles of $\Delta P(t)/2P_{M_0}$ as a function of $t$ according to Eq (15), where

$E_{appl}$=(500, 400, 300, 200, 150, and 100) kV/cm. The curves according to Eq (12)

($M_0$=1000) for these $E_{appl}$ have also been plotted for comparison.

Fig 4 (color) polarization switching data by Verdier *et al*. (Ref [23]) on commercialized PZT

bulk fitted by Eq (15).

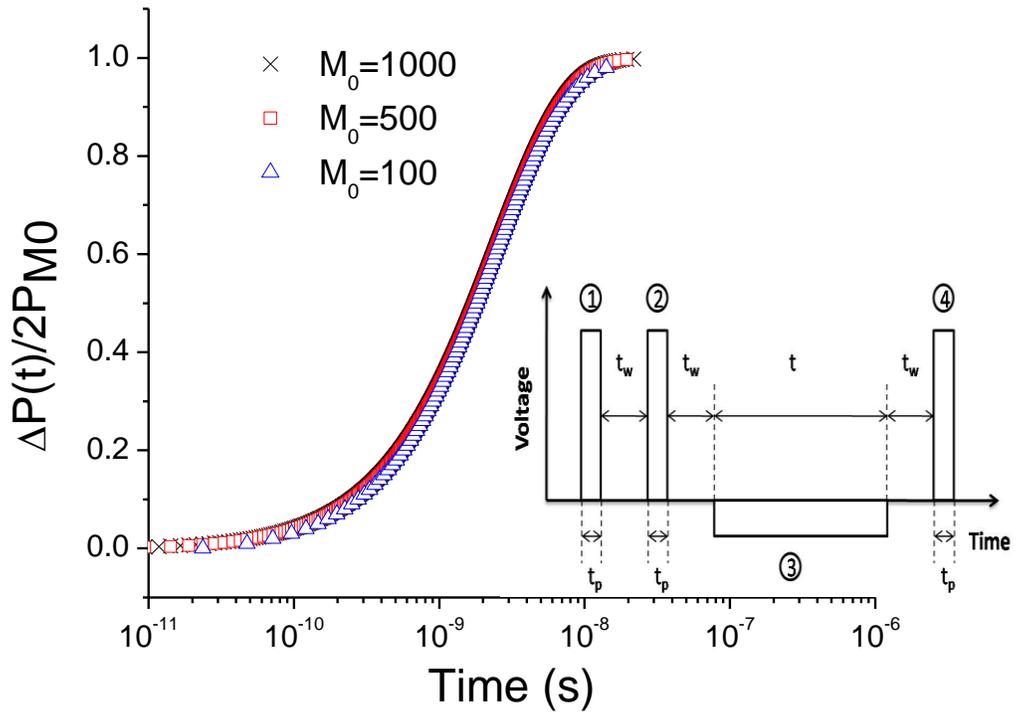

Fig 1

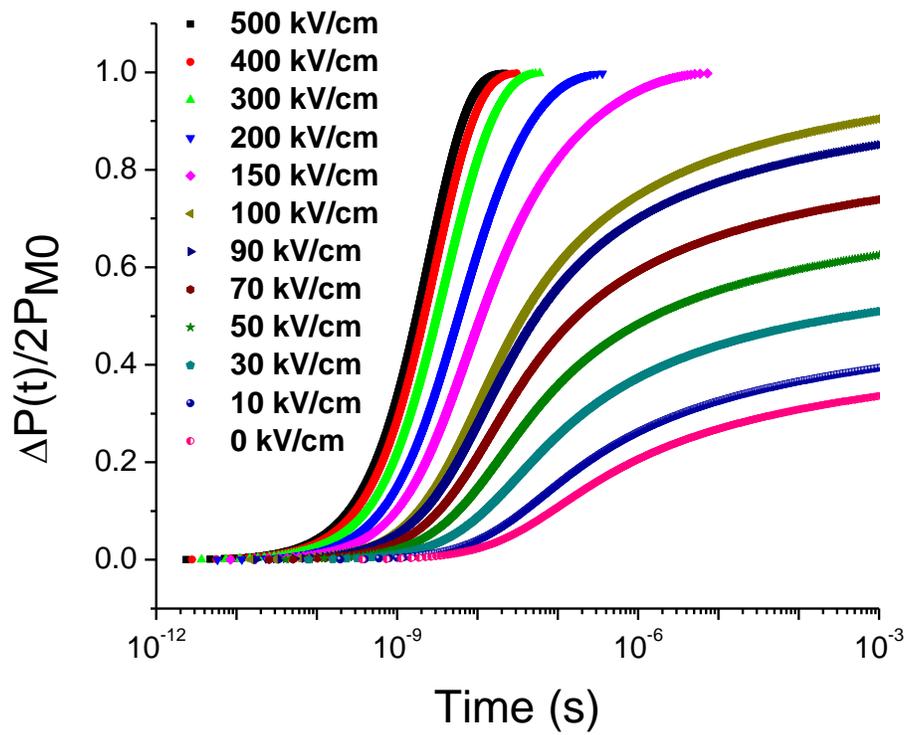

Fig 2

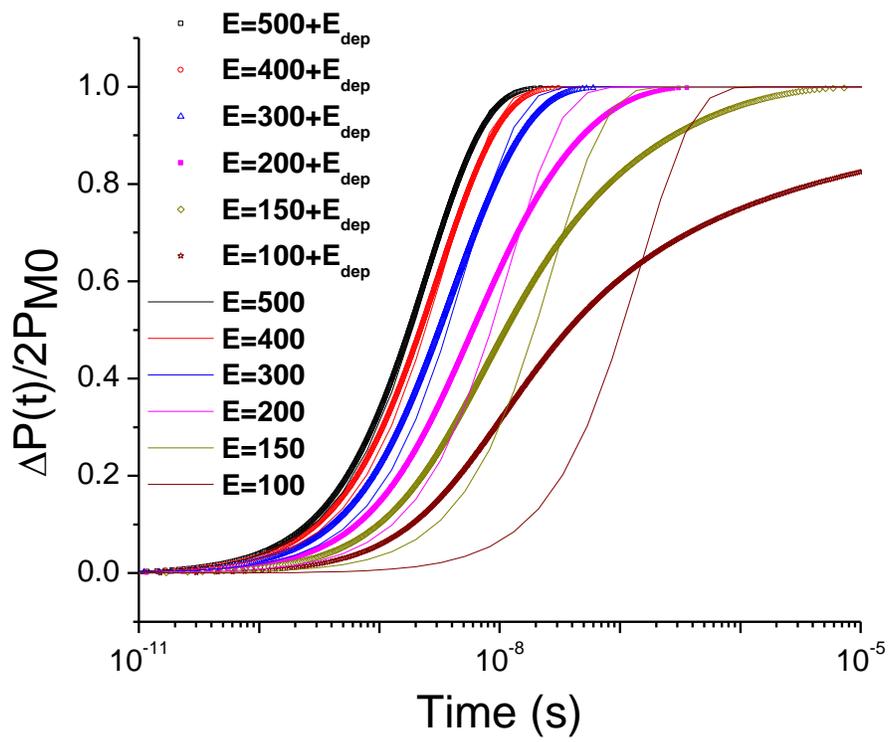

Fig 3

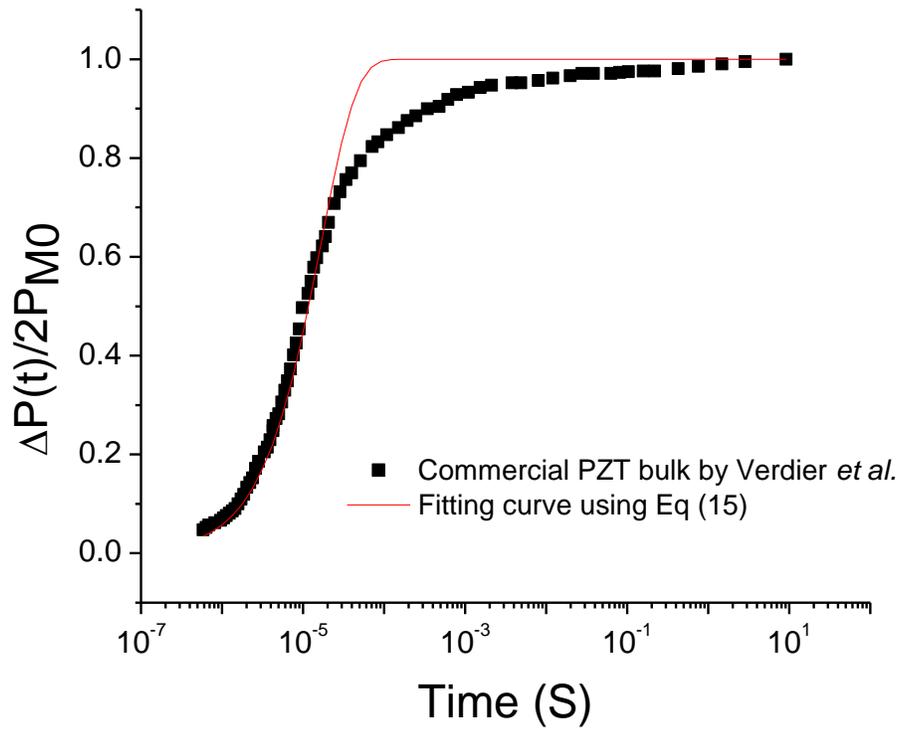

Fig 4